\documentclass[aps,prl,twocolumn,superscriptaddress]{revtex4}
\usepackage{amsmath}
\usepackage{graphicx}
\usepackage{dcolumn}
\usepackage{bm}
\usepackage{amsfonts}
\makeatletter
\def\captionof#1#2{{\def\@captype{#1}#2}}
\makeatother

\newcommand{\rme}{{\rm e}}
\newcommand{\rmi}{{\rm i}}

\newcommand{\Ud}[1]{\hspace{-0.5ex}\mathrm{d}{#1}\;}

\begin{document}
\title{Dielectric Breakdown of a Mott Insulator}

\author{Camille Aron }
\affiliation{Department of Physics and Astronomy, Rutgers University,
136 Frelinghuysen Rd., Piscataway, NJ 08854, USA}

\begin{abstract}
 We study the non-equilibrium steady state of a Mott insulator coupled to a thermostat and driven by a constant electric field, starting from weak fields, until the dielectric breakdown, and beyond. We find that the conventional Zener picture does not describe the steady-state physics. In particular, the current at weak field is found to be controlled by the dissipation. Moreover, in connection with the electric field driven dimensional crossover, we find that the dielectric breakdown occurs when the field strength is on the order of the Mott gap of the corresponding lower dimensional system. We also report a resonance and the melt-down of the quasi-particle peak when the field strength is half of this Mott gap.
\end{abstract}

\maketitle

An early achievement in the understanding of the non-linear response of electronic systems driven by strong electric fields is due to Zener in 1934~\cite{Zener}. He computed the rate of interband transitions of a one-dimensional non-interacting band insulator in a constant electric field, assuming that there is no back-feeding from the conduction to the valence band. This predicted a threshold electric field $E_{\rm th}$ above which the dielectric breakdown of the insulator occurs.

Following Oka, Arita, and Aoki's proposal that this single electron picture also applies to Mott insulators~\cite{OkaAritaAoki2003}, many efforts have been done to check their idea by testing Zener's formula: $E_{\rm th}\propto \Delta^2$ where $\Delta$ is the gap of the insulator.
Numerically, this out-of-equilibrium strongly interacting problem has been tackled by means of time-dependent (TD) methods such as TD density matrix renormalization group~\cite{TdDMRG} or TD exact diagonalization in $1d$ finite systems~\cite{TdED}, and  by TD dynamical mean-field theory in infinite dimensions~\cite{TD-DMFT}. 
There, the lack of a dissipation mechanism (necessary to get a non-trivial steady-state as earlier understood by~\cite{FukuiKawakami1998,SugimotoOnodaNagaosa2008}), causes a continuous heating up of the system~\cite{Amaricci}.

Experimentally, the electric field dependence of the current density is extracted from the current-voltage characteristic when applying a bias voltage on large samples~\cite{Natelson}. Several examples exhibit a much smaller threshold field than the estimation from Zener's formula~\cite{Yammanouchi1999}. 
In this Letter, we address this problem by driving out of equilibrium a two-dimensional ($2d$) Hubbard model coupled to a dissipative thermostat. We treat both the strong electric field and the strong interaction, and we bypass the transient dynamics by means of the non-equilibrium steady-state dynamical mean-field theory (NESS-DMFT) developed recently by the author and collaborators~\cite{AronKotliarWeber2012}. 

Hereafter, we describe the model and detail the computations. 
Then, we summarize the influence of the dissipation on the \textit{equilibrium} physics of the Mott transition.
Later, we study the influence of the electric field on the spectral properties of the Mott insulator and argue that the dissipation is the leading mechanism for the interband current.
Afterwards, we undertake the systematic exploration of the non-linear response of the system as the electric field is increased and as the dimensional crossover to the corresponding $1d$ system takes place, until the full dimensional reduction predicted on general grounds in~\cite{AronKotliarWeber2012}. 
In particular, we discuss a small jump in the conductivity and the melt-down of the quasi-particle peak when the field strength is half of the Mott gap of this $1d$ system. We also detail the physics of the dielectric breakdown that is found when the field strength is on the order of this Mott gap, contrary to Zener's picture.

\paragraph{Model.}
We consider the Hubbard model on a $d=2$ square lattice. The static and uniform electric field is set along an axis of the lattice: $\mathbf{E} = E \mathbf{u}_x$ with $E>0$.
The Lagrangian of the system coupled to its environment reads (we set $\hbar=1$ and use the conventions of~\cite{AronKotliarWeber2012})
\vspace{-0.5em}
\begin{equation}
\begin{array}{rl}\label{eq:Lagrangian}
\mathcal{L} =& \displaystyle \sum_{i \sigma} \bar c_{i \sigma} \left[ \rmi\partial_t -  \phi_{i}(t) \right] c_{i \sigma} 
 - U \sum_{i}  \bar c_{i \uparrow}  c_{i  \uparrow}  \bar c_{i \downarrow}  c_{i \downarrow} \\ 
  & +  \displaystyle \sum_{\langle i j \rangle \sigma}   \bar c_{i \sigma} t_{ij} \rme^{\rmi\alpha_{ij}(t)} c_{j \sigma} 
+ \mathrm{conj.} \\
& \quad + \ \displaystyle \gamma \sum_{i \sigma l} \rme^{\rmi\theta_i(t)} \bar b_{i\sigma l} c_{i\sigma} + \mbox{conj.} \;,
\end{array}
\vspace{-0.5em}
\end{equation}
where $c_{i\sigma}$ and $\bar c_{i\sigma}$ are the Grassmann fields representing an electron at site $i$ with spin $\sigma\in\{\uparrow,\downarrow\}$.
$U$ is the on-site Coulombic interaction and $t_{ij} \equiv (a/2\pi)^{2} \int\Ud{\mathbf{k}} \rme^{\rmi\mathbf{k}\cdot\mathbf{x}_{ij}} \epsilon(\mathbf{k}) $ sets the hopping amplitude between two nearest neighbors distant of $a$: $\epsilon(\mathbf{k}) = \epsilon_0 \left[\cos(k_x a)+\cos(k_y a) \right]$, each dimension contributing by $2\epsilon_0$ to the bandwidth of the equilibrium non-interacting ($E=U=0$) system.
Integrals over $k_x$ and $k_y$ are computed  between $-\pi/a$ and $\pi/a$.
The last term in (\ref{eq:Lagrangian}) is the coupling to the thermostat which is composed of independent non-interacting electronic reservoirs in equilibrium at a very low temperature $T$ and a chemical potential $\mu_0=U/2$ in order to work at half-filling \textit{i.e.} with one electron per site in average (we also restrict ourselves to the paramagnetic solution and drop the spin indices). $\gamma$ is a real hopping parameter, the $b$'s represent the electrons in the reservoirs, and $l$ labels their energy levels.
The Peierls phase factors, $\alpha_{ij}(t) \equiv q \int_{\mathbf{x}_j}^{\mathbf{x}_i} \Ud{\mathbf{x}} \cdot \mathbf{A}(t,\mathbf{x})$ and  $\theta_i(t) \equiv  \int^t \Ud{t'} \phi_{i}(t')$, are required by the gauged $\mathrm{U}(1)$ symmetry associated with the conservation of the charge $q$ of the electrons. $\phi$ and $\mathbf{A}$ are the scalar and vector potentials: $\mathbf{E} = -\boldsymbol{\nabla} \phi - \partial_t \mathbf{A}$. $|q|Ea$ is the energy an electron acquires when hopping to a neighboring site under the work of the electric field.
To work with gauge-invariant quantities, we use the variables $\varpi \equiv \omega - \phi$ and $\boldsymbol{\kappa}\equiv \mathbf{k} - q \mathbf{A}$ and later absorb the Hartree shift by redefining  $\varpi-U/2$ into $\varpi$. 

An efficient dissipation is achieved if the bandwidth $W$ of the local density of states (DOS) of the reservoirs is the largest energy scale. The other details of these DOS are not relevant and we take them to be Gaussian, yielding a contribution of the dissipation to the Keldysh self-energy: $\Sigma_{th}^K(\varpi) = \Gamma \exp(-\varpi^2/\pi W^2) \tanh(\varpi/2k_{\rm B}T)$ where $\Gamma\equiv\gamma^2/W$. We work at small dissipation $\Gamma$ but large enough 
for the momentum resolved spectral function to be positive everywhere, ensuring a stable steady state. Otherwise, this signals oscillatory instabilities (such as Bloch oscillations) developing on top of the steady-state solution~\cite{DaviesWilkins1988}.

\paragraph{Computational details.}
The non-equilibrium steady state is solved by means of the NESS-DMFT algorithm developed in~\cite{AronKotliarWeber2012} and based on a gauge-invariant Schwinger-Keldysh formalism~\cite{OnodaSugimotoNagaosa2006}. The interaction contribution to the retarded and Keldysh self-energies ($\Sigma_U^R$ and $\Sigma_U^K$) are computed using second order iterated perturbation theory in $U$ (IPT) as the impurity solver. Although it is not a $\Phi$-derivable approximation, it is a current conserving approximation at half-filling~\cite{HershfieldDaviesWilkins1992}.
For each value of the electric field, the dressed retarded and Keldysh Green's functions ($G_U^R$ and $G_U^K$) are obtained in the strongly interacting regime by starting from the non-interacting solution, then by slowly increasing the interaction ($U\mapsto U+\delta U$) while converging at each step the impurity and the following lattice equations [see Eqs.~(7), (8), (9), and (3) in \cite{AronKotliarWeber2012}],
\begin{eqnarray}
 G_U^R &=& G_{U-\delta U}^R + G_{U-\delta U}^R \ast \delta \Sigma^R \ast G_{U}^R \;, \\
   G_{U}^K &=& G_{U}^R \ast \Sigma^K  \ast {G_{U}^{R}}^*\;, \label{GK}
\end{eqnarray}
where $\delta \Sigma^R \equiv \Sigma^R_{U}  - \Sigma^R_{U-\delta U}$ and $\Sigma^K \equiv \Sigma_{th}^K + \Sigma_U^K$.
To take further advantage of both the mean-field approximation and the geometry of the setup, the star product is evaluated in the mixed $(\varpi;n_x;\kappa_y)$-space where $n_x\in\mathbb{Z}$:
\begin{align}\label{eq:MoyalP} 
\left[ f \ast g\right] (\varpi;n_x;\kappa_y)\!=& \!  \sum_{m_x} f\left(\varpi+m_x{q{E}a}/2;n_x-m_x;\kappa_y \right) \nonumber \\
  & \hspace{-2em} \times \displaystyle g\left(\varpi+(m_x-n_x){q{E}a}/2;m_x;\kappa_y \right), 
\end{align}
with $f(\varpi;n_x;\kappa_y)\equiv (a/2\pi) \int\Ud{\kappa_x} \rme^{\rmi \kappa_x n_x a}f(\varpi,\boldsymbol{\kappa})$. Each evaluation of Eqs.~(2) and (3) requires performing a single numerical summation and the overall computation is slower than the equilibrium algorithm by a factor $2N_x = 2\times2\pi/a\delta \kappa_x$ only, where $\delta \kappa_x$ is the discretization step for $\kappa_x$. Hereafter, numerical results are obtained with $\epsilon_0 = a=q=k_{\rm B}=1$.

\begin{figure}[t]
\centerline{
\includegraphics[width=8.5cm]{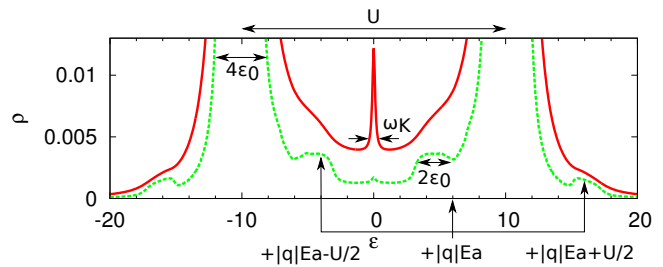}
\vspace{-0.7em}
}
\caption{\label{fig:islands}\label{fig:DOS}  \footnotesize In-gap DOS for $U=20$ and $E=6$ ($T=0.05$). Dotted line: very weak dissipation $\Gamma=0.09$ (case presenting oscillatory instabilities) revealing the BZ ``archipelagos'' centered around $\epsilon_{\rm a}=\pm qEa$ and both composed of two $1d$ BZ islands at $\epsilon_{\rm a}\pm U/2$.  Solid line: the same for $\Gamma=0.25$ where the quasi-particle peak around $\epsilon=0$ is much stronger and the details of the BZ islands are now almost indistinguishable.
}
\end{figure}

\begin{figure}[t]
 \centerline{
 \includegraphics[height=3.1cm]{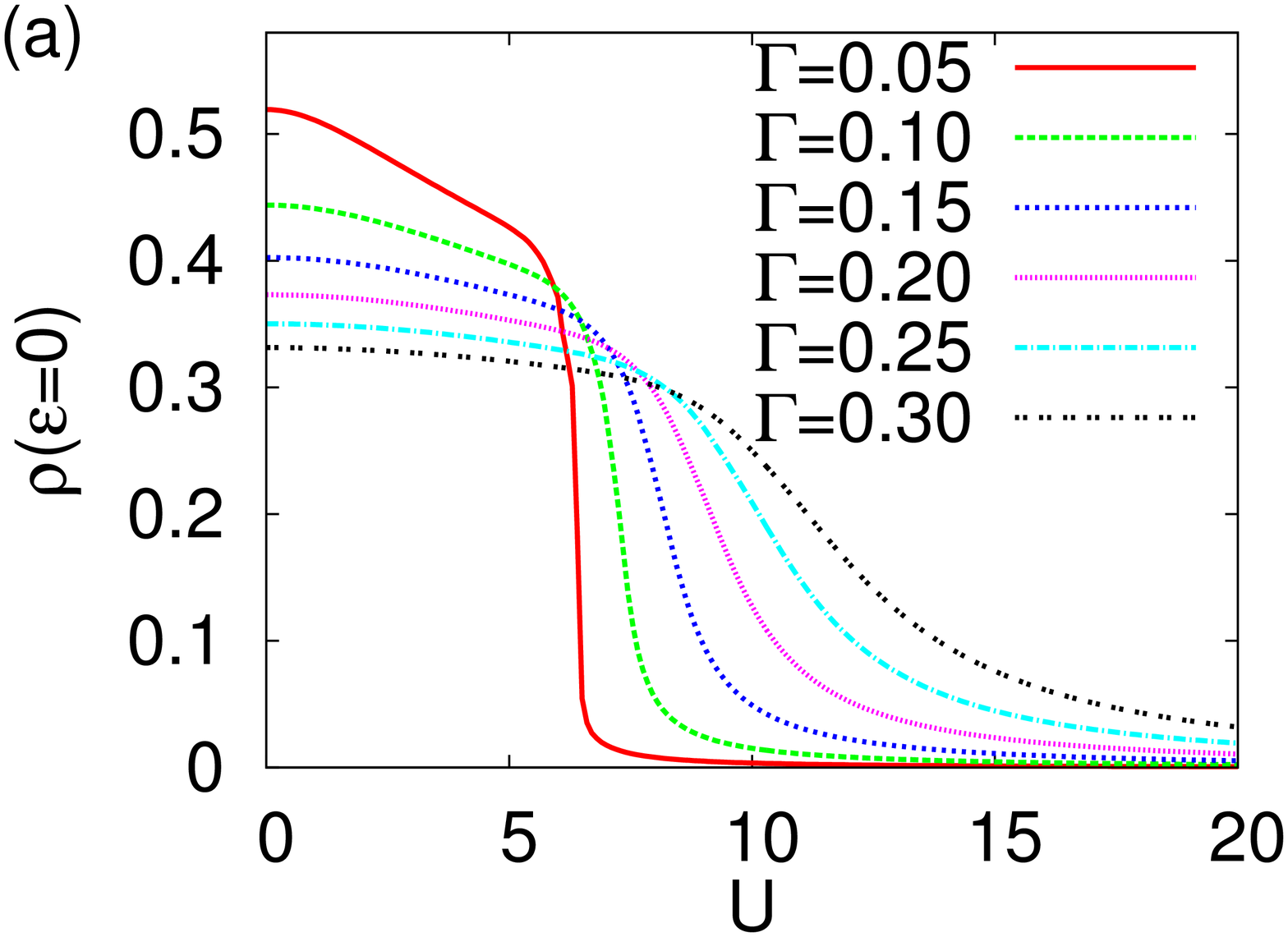}
 \includegraphics[height=3.1cm]{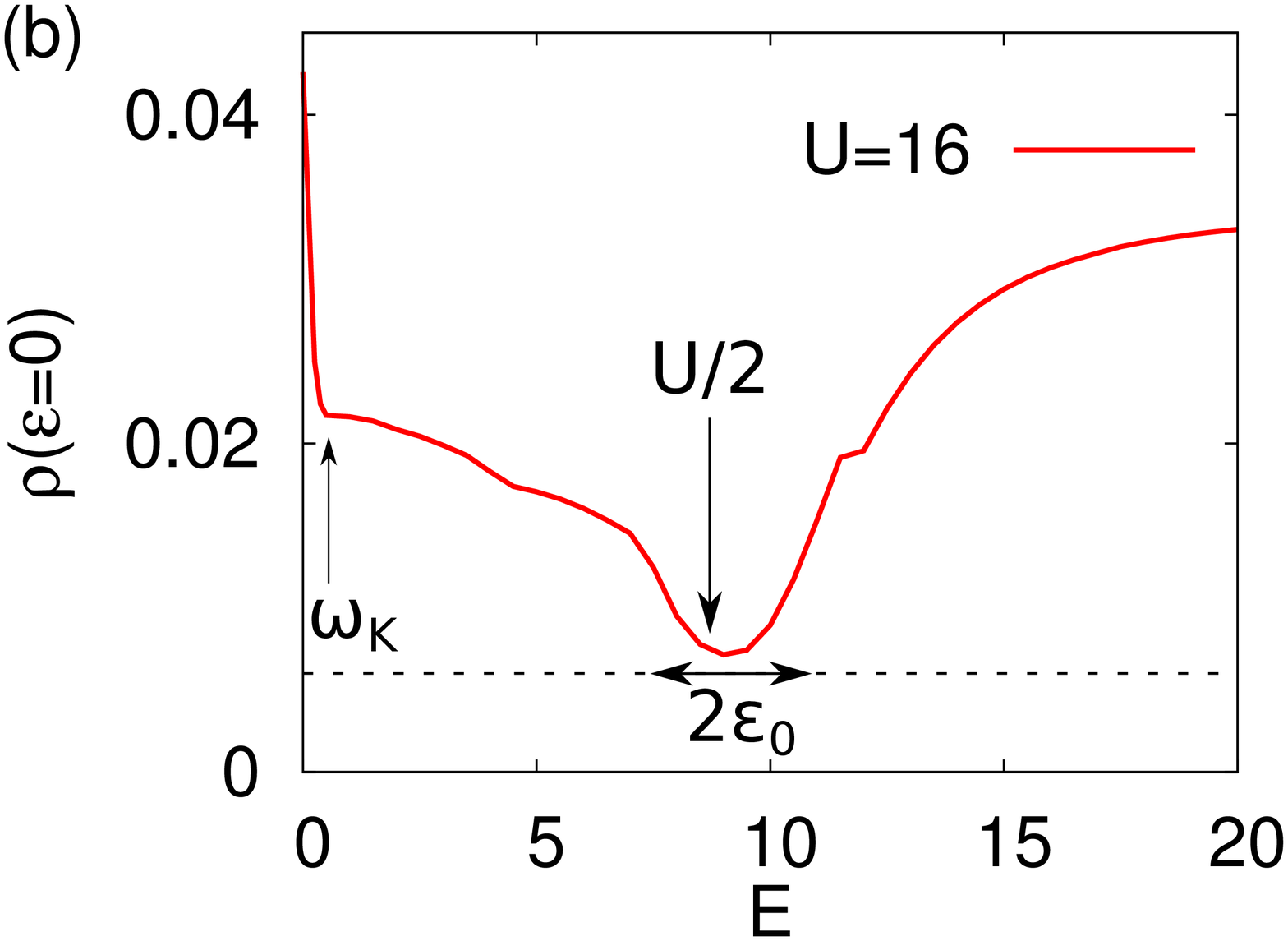}
 \vspace{-0.5em}
}
\caption{\label{fig:tinyJ} \footnotesize (color online) (a) Height of the equilibrium quasi-particle peak, $\rho(\epsilon=0)$, as a function of $U$ for different dissipations $\Gamma$ ($E=0$, $T=0.05$). (b) The two melt-downs of the quasi-particle peak at $E\lesssim\omega_K$ and $E\simeq U/2$, followed by the growth of the peak of the equilibrium $1d$ model. Dashed line: level of the dissipative background at $\epsilon\simeq0$ estimated from the equilibrium data ($U=16, T=0.05,\Gamma=0.25$).
}
\end{figure}

\paragraph{Influence of the dissipation in equilibrium.}
In equilibrium ($E=0$) and as the interaction $U$ is increased, the Hubbard model exhibits a well-known quantum phase transition from a metal to a Mott insulator  characterized by the opening of an energy gap $\Delta \simeq U - 2d\epsilon_0$ separating the so-called Hubbard bands~\cite{BrinkmanRice1970}.
The presence of a weak dissipation $\Gamma$ smoothens the sharp features of the spectral function over an energy window $\Gamma$. In particular, the edges of the Hubbard bands leak into the gap, responsible for a dissipative in-gap DOS controlled by $\Gamma/U^2$. 
Dissipation also delays the transition which turns into a smooth crossover taking place onto an extended region in $U$~\footnote{The metastable region between $U_{c1}$ and $U_{c2}$ closes in presence of a weak dissipation.}. There, what is left of the metal manifests itself by a weakly dispersive Kondo-like resonance of width $\omega_K$, centered around the Fermi level, and containing a fraction $Z$ of all the states.
When increasing $U$, the height of this quasi-particle peak is first roughly constant (and decreases with $\Gamma$) while $\omega_K$ decreases continuously. Deep in the strongly interacting phase, the peak becomes controlled by the dissipation as $\omega_K$ is rather constant (and set by $\Gamma$) while its height vanishes as $1/U^2$ (and grows with $\Gamma$) [see its dependence on $U$ and $\Gamma$ in Fig.~\ref{fig:tinyJ}(a)].

\begin{figure}[t]
\centerline{
\includegraphics[height=3.2cm]{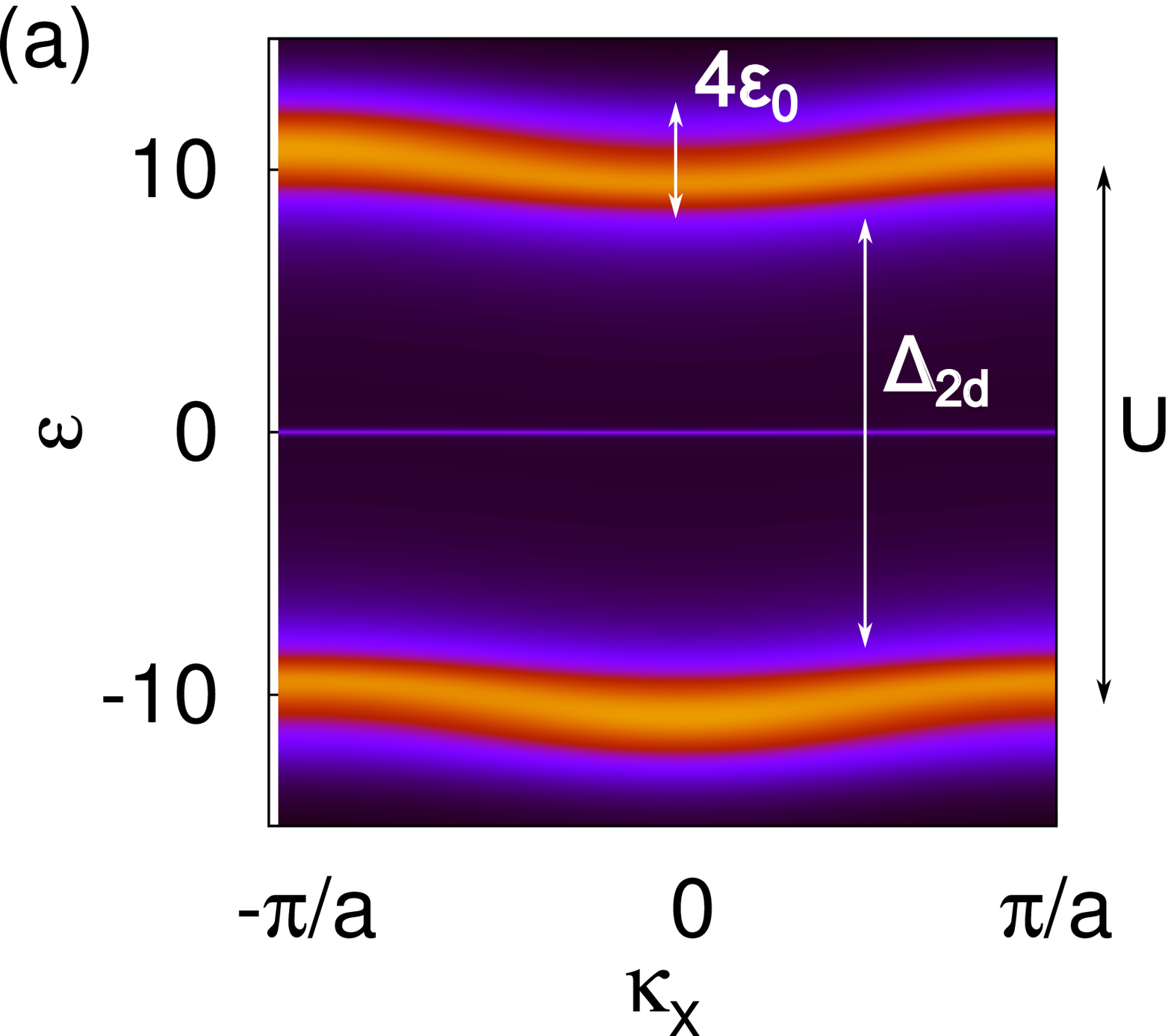}
\includegraphics[height=3.2cm]{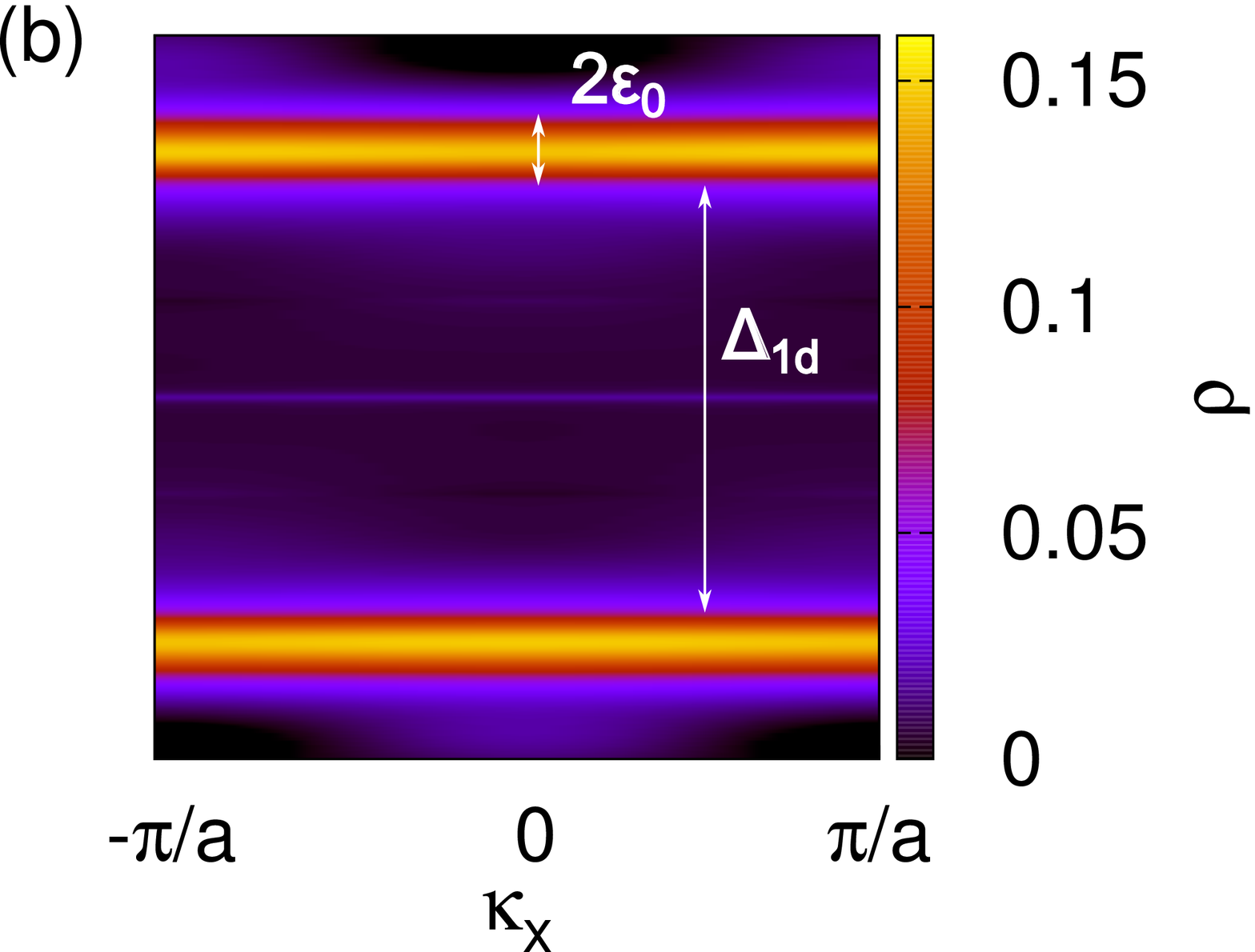}
\vspace{-1.em}
}
\caption{\label{fig:spectral1d} \footnotesize (color online) 
(a) Equilibrium spectral function $\rho(\epsilon,\boldsymbol{\kappa})$ integrated over $\kappa_y$ for $U=20$.
(b) The same for $2\epsilon_0=2 < E=8 < U=20$ is now almost $\kappa_x$-independent and the bands have a width $2\epsilon_0$, similarly to the 1$d$ model. ($T=0.05, \Gamma=0.35$).}
\end{figure}

\paragraph{Influence of the electric field on the spectral properties.}
Since deep in the strongly interacting regime, each Hubbard band exhibits the spectral features of a single non-interacting band and $U$ only enters through the gap $\Delta$ [see Fig.~\ref{fig:spectral1d}(a)], one can expect the effects of the electric field on the spectral function of the Mott insulator to be similar to the case of a non-interacting band insulator. \\
In the well-known case of a single non-interacting band, some Bloch-Zener (BZ) islands appear in the DOS beyond the edges of the band, equally spaced in energy by $|q|Ea$, with a weight that is \textit{exponentially} killed on a scale $\epsilon_0^{1/3} (|q|Ea)^{2/3}$ as one gets away from the band edges, and the energy structure of which is controlled by the DOS of the \textit{equilibrium} $1d$ system along the $y$-direction~\cite{DaviesWilkins1988}. \\
In our Mott insulator case, we find that this scenario indeed occurs as we observe similar islands in the DOS. They have the structure of the corresponding equilibrium 1$d$ Mott insulator. Since the latter has a gapped DOS, the islands are in fact ``archipelagos'' centered on multiples of $\pm qEa$ and composed of two islands of width $2\epsilon_0$ and separated by $U$. We illustrate in Fig.~\ref{fig:islands} this rich structure of the DOS between the Hubbard bands. These in-gap islands allow the transition of carriers from the lower to the upper Hubbard band by successive excitations of energy $|q|Ea$.
However, the dissipation creates a continuous in-gap DOS, damped as a \textit{power law} as one gets away from the band edges, and it is therefore expected to be the main contribution for those in-gap states (see Fig.~\ref{fig:islands}). It was indeed the case in all the stable steady states we expolred.

Before we start the systematic study of the non-linear regime, notice that as the field is increased, the system experiences a dimensional crossover from the insulating phase of the 2$d$ equilibrium Hubbard model (at $E=0$) to the insulating phase of the 1$d$ equilibrium model (when $|q|Ea$ is the largest energy scale)~\cite{AronKotliarWeber2012}. Since both exhibit a similar DOS deep in the strongly interacting regime (at least for the paramagnetic solutions obtained with the local approximation of the single-site DMFT), one expects a smooth variation from the $2d$ DOS with a pseudo gap $\Delta_{2d}\simeq U-4\epsilon_0$ between bands of width $4\epsilon_0$ towards the $1d$ DOS with a pseudo gap $\Delta_{1d}\simeq U-2\epsilon_0$ between bands of width $2\epsilon_0$~\cite{DMFT}. Therefore, the qualitative features of the current characteristic can already be predicted by simply reasoning on Fig.~\ref{fig:DOS}.

\vspace{1em}

\begin{figure}[t]
\centerline{
\includegraphics[height=4.8cm,angle=-90]{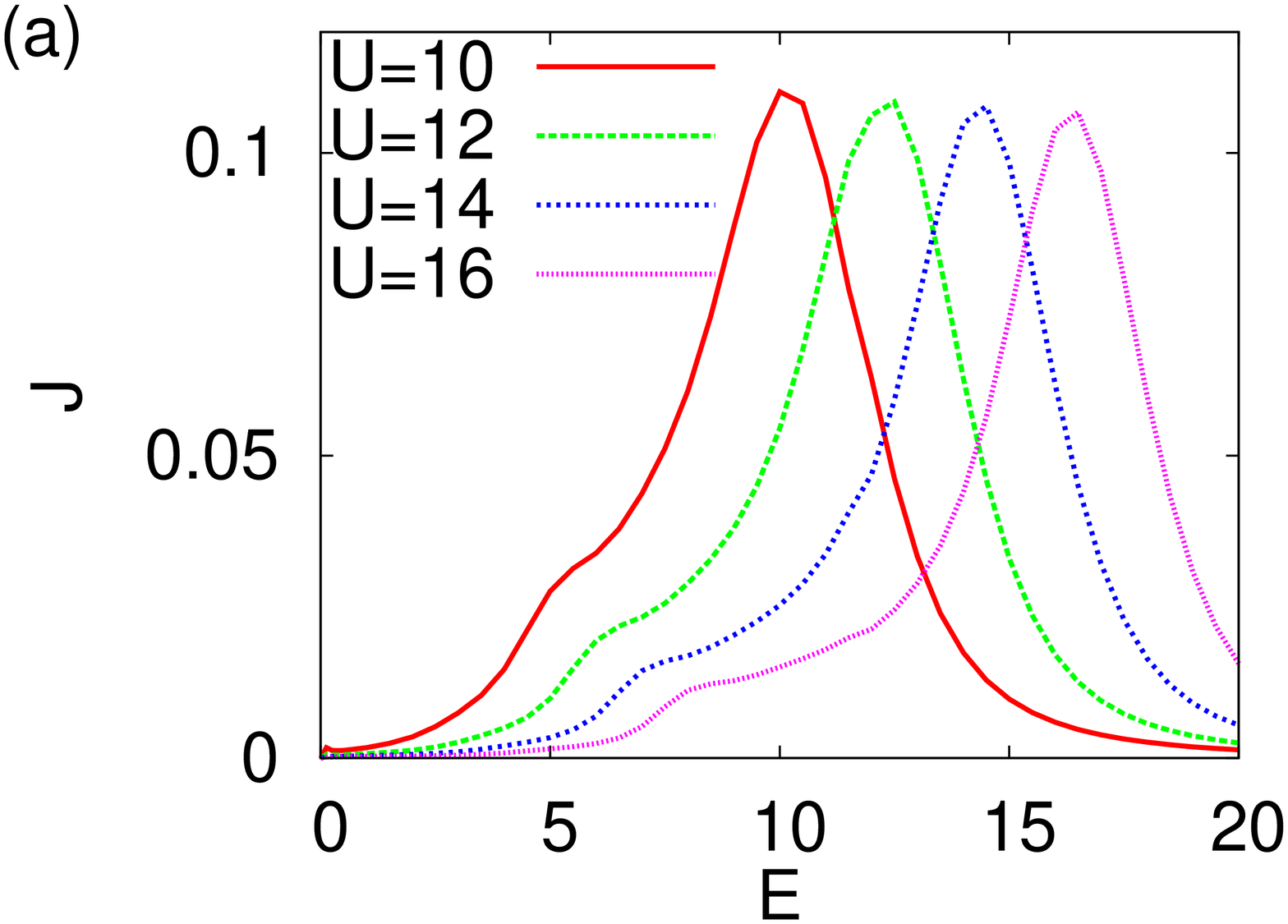}
\hspace{-2.em}
\includegraphics[height=4.8cm,angle=-90]{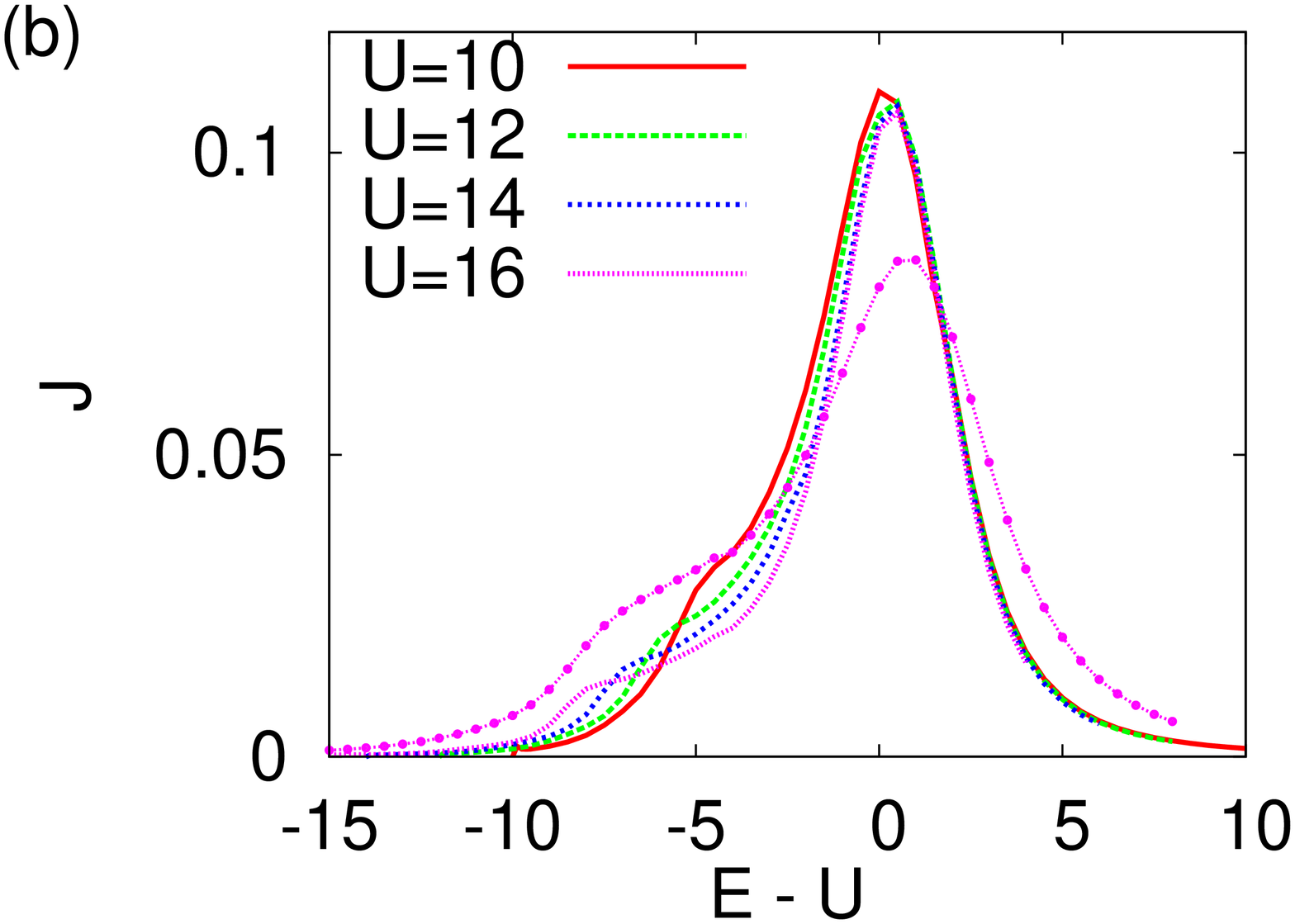}
\vspace{-0.5em}
}
\caption{\label{fig:current} \footnotesize (color online) (a) Current density $J(E)$ for different $U$ ($T=0.05, \Gamma=0.25$). The first jump is located at $E \simeq U/2$ and the maximum at $E \simeq U$. (b) The same data is plotted against $E-U$ to prove the scaling $J(E-U)$ in the metalized regime. The dotted curve corresponds to $\Gamma=0.50$.}
\end{figure}

Below, we detail the fate of the insulating phase when increasing the electric field by focusing on the momentum resolved spectral function $\rho(\epsilon,\boldsymbol{\kappa})$, the local DOS $\rho(\epsilon)$, and the current density $J(E)$  plotted in Fig.~\ref{fig:current}.
The latter also provides qualitative informations on the asymmetry in $\kappa_x$ of the momentum distribution function $n(\boldsymbol{\kappa})$: $J \propto -2 \int\Ud{\boldsymbol{\kappa}} \partial_{\kappa_x} \epsilon(\boldsymbol{\kappa}) \ n(\boldsymbol{\kappa})$.

\paragraph{$|q|Ea \ll 2\epsilon_0 \ll U$.} Let us start with very weak fields. At moderate values of $U$ for which the quasi-particle peak is still present ($Z>0$), a small electric field such as $|q|Ea \lesssim  \omega_K$ can excite the states lying in an energy shell $\omega_K$ below the Fermi level ($\epsilon=0$) to the empty states above, resulting in a tiny current.
This reorganization of the distribution of occupied states around $\epsilon=0$ is qualitatively similar to having an effective temperature, and causes the partial melt-down of the quasi-particle peak [see Fig.~\ref{fig:tinyJ}(b)]~\cite{HershfieldDaviesWilkins1991}. 
As soon as  $|q|Ea$ is larger than $\omega_K$, the transition rate is now controlled by the small in-gap DOS created by the leakage of the Hubbard bands, leading to a drop in the differential conductivity.
Deep in the strongly interacting regime, the quasi-particle peak vanishes ($Z\to0$) and the growth of the tiny current is controlled by the dissipative in-gap DOS which is on the order of $\Gamma /\Delta^2$.

\begin{figure}[t]
\centerline{
\includegraphics[height=5.3cm,angle=-90]{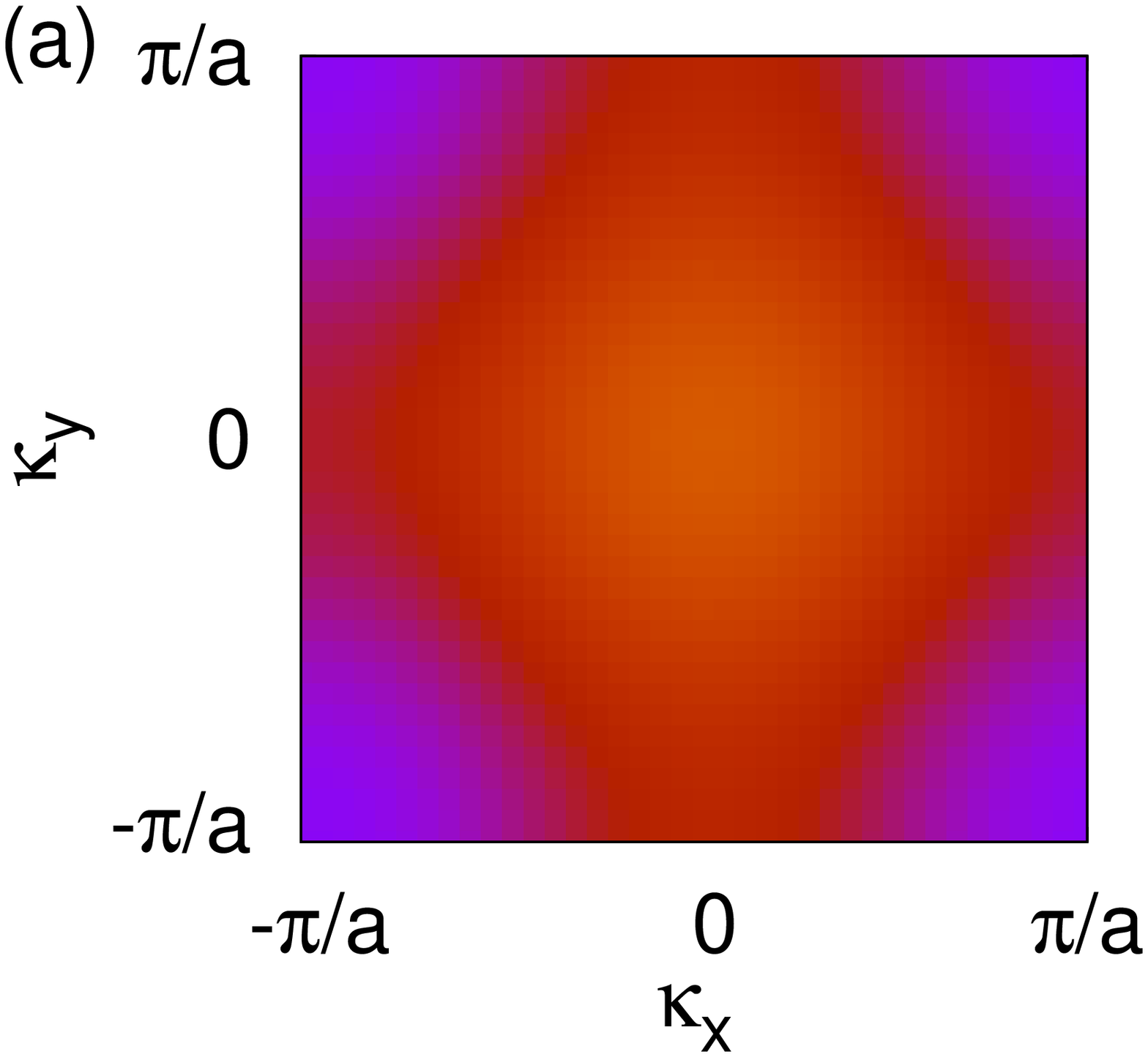}
\hspace{-3.em}
\includegraphics[height=5.3cm,angle=-90]{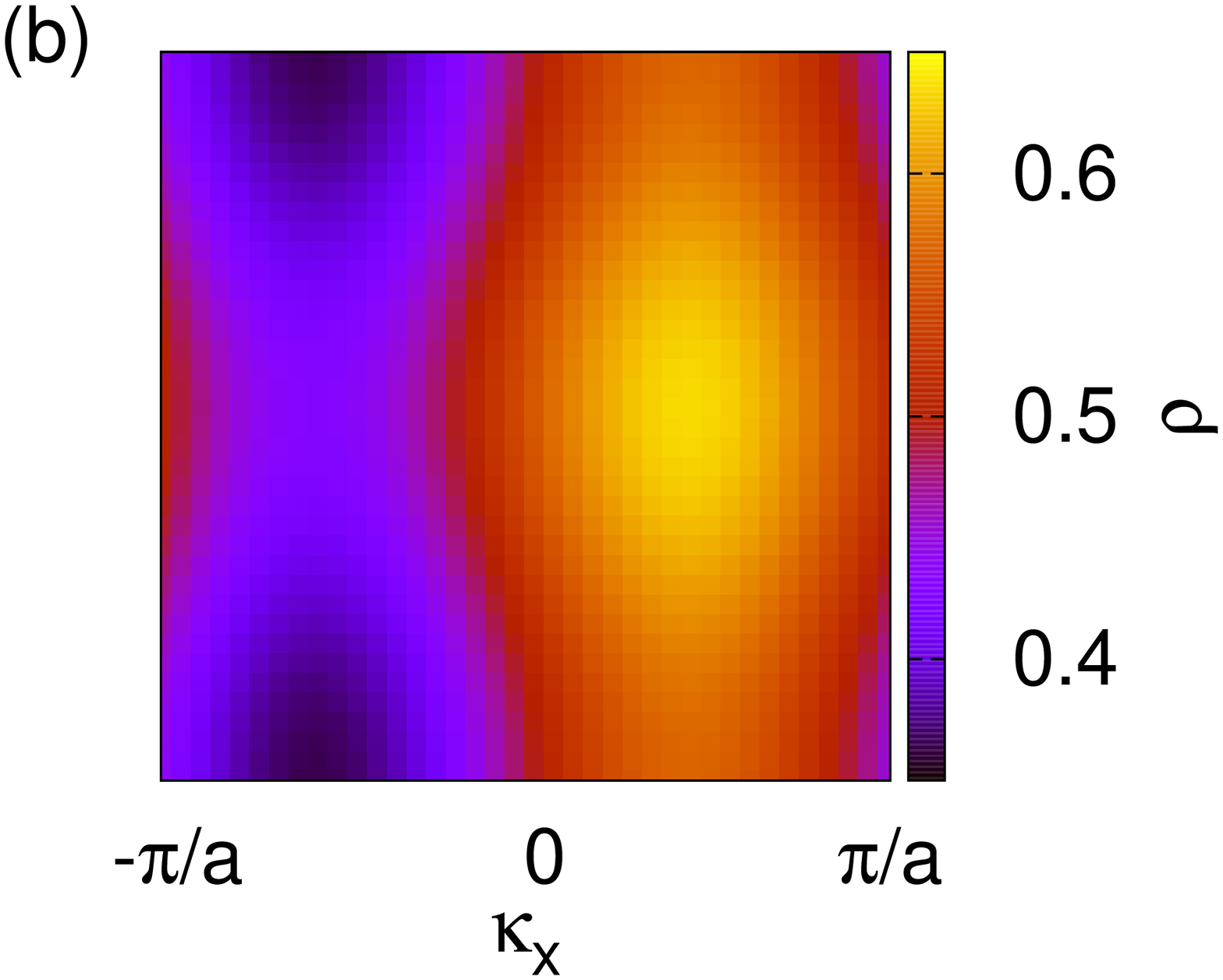}
\vspace{-1.em}
}
\caption{\label{fig:number} \footnotesize (color online) (a) The momentum distribution function $n(\boldsymbol{\kappa})$ for $2\epsilon_0=2 < E=8 < U=20$ is similar to $2d$ equilibrium [contrary to $\rho(\epsilon,\boldsymbol{\kappa})$ in Fig.~\ref{fig:spectral1d}(b)]. (b) $n(\boldsymbol{\kappa})$ just after the dielectric breakdown for $2\epsilon_0=2 < E=20 \simeq U=20$. ($T=0.05, \Gamma=0.35$).}
\end{figure}

\paragraph{$2\epsilon_0 \ll |q|Ea \ll U$.}
As the electric field intensity gets larger $2\epsilon_0$ (\textit{i.e.} the fraction of the non-interacting bandwidth corresponding to the $x$ direction), 
$\rho(\epsilon,\boldsymbol{\kappa})$ loses much of its dependence on $\kappa_x$ and becomes essentially the one of the $1d$ Hubbard model [see Fig.~\ref{fig:spectral1d}(b)]~\footnote{In the regime $\epsilon_0 \ll E \ll U$,  $\rho(\epsilon,\boldsymbol{\kappa})$ still weakly depends on $\kappa_x$ to recover a $\kappa_x$-dependent $n(\boldsymbol{\kappa})$ from Eq.~(\ref{GK}).}.
This is a first step towards the full dimensional reduction of the system. Meanwhile, 
$n(\boldsymbol{\kappa})$ is still very close to the one of the 2$d$ Hubbard model in equilibrium [see Fig.~\ref{fig:number}(a)] and the current is very weak.

\paragraph{$2\epsilon_0 \ll |q|Ea \sim U/2$.}
When the electric field intensity is comparable with the energy separating the lower Hubbard band with the Fermi level ($\epsilon=0$), $|q|Ea \simeq \Delta_{1d}/2 \simeq U/2 - \epsilon_0$, carriers can be excited from the former to the dissipative background around the latter. Concomitantly, the large amount of vacant states offered by the upper band favors a rapid pumping of these newly occupied states to the upper band. These resonant processes contribute
to a significant increase of the current density until $|q|Ea \simeq U/2$. Here again, a stronger dissipation favors a larger current \textit{via} the increase of the in-gap DOS. Notice also that the combination of the BZ archipelagos centered at $\pm qEa$ creates a large mid-gap BZ island on top of the dissipative background (see Fig.~\ref{fig:islands}) which also contributes to this resonance.
Furthermore, the reorganization of the distribution of occupied states around $\epsilon=0$ (the fraction of occupied states decreases significantly just below $\epsilon=0$ while it increases symmetrically above) is qualitatively similar to having a high effective temperature. This explains the complete melt-down of the quasi-particle peak that we observe until the resonance is broken when $|q|Ea \gtrsim U/2 + \epsilon_0$ [see Fig.~\ref{fig:tinyJ}(b)].

\paragraph{$2\epsilon_0 \ll |q|Ea \sim U$.}
When the electric field is of the magnitude of the Mott gap,  $|q|Ea \simeq \Delta_{1d}$, carriers in the lower band can directly populate the upper band. This is the dielectric breakdown.
The electric current increases rapidly as the field is further increased and reaches a maximum at $|q|Ea \simeq U$.
After the dielectric breakdown, one expects $U$ to be quite irrelevant and the scaling $J = J(E-U)$ to hold [see Fig.~\ref{fig:current}(b)] since the structure and the occupation of each Hubbard band are almost independent of $U$, except for small corrections due to the dissipative background and the quasi-particle peak (if any). The Hubbard bands can be seen as two non-interacting systems connected to a thermostat: electrons are excited from the first one to the second, then are absorbed by the thermostat which also repopulates the first system.
The dissipation enters this picture in two ways. One the one hand, a stronger dissipation accelerates the repopulatation of the lower Hubbard band and should therefore favor a larger current. On the  other hand, the dissipation is expected to reduce the current because it is responsible for fewer states in the Hubbard bands since they leak into the gap and since it also strengthens the quasi-particle peak. All together, we show in Fig.~\ref{fig:current}(b) that a stronger dissipation favors a smaller value of the maximum current but a larger current away from this maximum. 
Together with the sharp current increase, the weight of 
$n(\mathbf{\kappa})$
is strongly displaced along $\kappa_x$ [see Fig.~\ref{fig:number}(b)]. Notice the sharper discontinuity of 
$n(\mathbf{\kappa})$ due to the fact that the lower (upper) Hubbard band has now a sizable fraction of unoccupied (occupied) states.

\paragraph{$2\epsilon_0 \ll  U \ll |q|Ea$.}
When the electric field is stronger than any other energy scale, the dimensional reduction predicts that the system behaves as a collection of uncoupled $1d$ Hubbard chains in equilibrium~\cite{AronKotliarWeber2012}. The DOS being bounded, the electric field is too strong for any transition to take place (except in the outer dissipative background) as soon as $|q|Ea \gtrsim U + 2\epsilon_0$. Both $\rho(\epsilon,\boldsymbol{\kappa})$ and $n(\boldsymbol{\kappa})$ are the ones of the $1d$ Hubbard model in equilibrium and the current vanishes accordingly.

\paragraph{Discussion.}
We have investigated the steady-state physics of a $2d$ Mott insulator driven out of equilibrium by a constant electric field and coupled to a thermostat.
We argued that the interband current is mostly due to the presence of in-gap states created by the dissipation.
Also contrary to Zener's picture, we observed the dielectric breakdown of the Mott insulator after $|q|Ea\simeq \Delta_{1d}$. Furthermore, we revealed a resonance around $|q|Ea\simeq U/2$ responsible for a small jump in the conductivity and for the melt-down of the quasi-particle peak.
We also showed that the dimensional crossover takes place on two separated energy scales: the spectral properties turn to the ones of the $1d$ Mott insulator as soon as $|q|Ea\gg2\epsilon_0$, whereas the distribution functions only reach thermal equilibrium in $1d$ when $|q|Ea \gg U$.

We expect this scenario to be also relevant for $3d$ samples crossing over to $2d$, where the DMFT solutions are all the more valid.
We also believe that our results can be put to experimental test with cold atoms trapped in optical lattices where strong electric fields ($|q|Ea > U$) can be mimicked by forcing the lattice potential~\cite{cold-atoms} and the dissipation can be engineered by coupling the Mott insulator to a superfluid fraction of the atomic condensate.

I am grateful to A. Amaricci, P. Coleman, K. Haule, G. Kotliar, O. Parcollet, J. Simonet, and C. Weber for  discussions and comments. I would like to thank especially G. Kotliar for his stimulating guidance. This work has been supported by NSF grant No. DMR-0906943.

\end{document}